\documentclass[aps,prb,a4paper,superscriptaddress,twocolumn,citeautoscript,nofootinbib,nobalancelastpage,longbibliography,nobibnotes]{revtex4-1}

\makeatletter
\@booleanfalse\footinbib@sw
\makeatother

\usepackage[verbose=true]{microtype}

\usepackage{amsmath}
\usepackage{amssymb}
\usepackage{txfonts}        
\usepackage{quantum}

\usepackage[symbol,perpage,flushmargin]{footmisc}
\usepackage{paralist}
\usepackage{enumitem}
\usepackage{subfigure}
\usepackage{graphicx}

\usepackage{array}
\usepackage{multirow}
\usepackage{rotating}

\usepackage[colorlinks,linktocpage,hypertexnames=false,pdfpagelabels]{hyperref}
\usepackage[amsmath,thmmarks,hyperref]{ntheorem}
\usepackage[capitalise]{cleveref}

\usepackage{color}
\usepackage{epsfig}
\usepackage{symbols}

\newtheorem{theorem}{Theorem}

\theoremstyle{nonumberplain}
\newtheorem{main-result}[theorem]{Main Result}

\theoremstyle{break}

\theoremstyle{nonumberplain}
\theorembodyfont{\normalfont}
\theoremsymbol{$\Box$}

\theoremstyle{nonumberbreak}
\theorembodyfont{\normalfont}
\theoremsymbol{$\Box$}

\crefname{equation}{}{}
\crefrangelabelformat{equation}{\textup{(#3#1#4)}--\textup{(#5#2#6)}}
\crefname{term}{term}{terms}
\creflabelformat{term}{(#2#1#3)}
\crefrangelabelformat{term}{(#3#1#4)--(#5#2#6)}
\crefname{figure}{Figure}{Figure}
\crefname{remark}{Remark}{Remarks}
\crefname{obstacle}{obstacle}{obstacle}

  {\par\noindent\ignorespacesafterend}
  {\par\noindent\ignorespacesafterend}
  {\par\noindent\ignorespacesafterend}

\makeatletter
\renewcommand\section{%
  \@startsection{subsection}{1}{0pt}%
  {-\baselineskip}{.2\baselineskip}%
  {\normalfont\normalsize\bfseries\raggedright\color{blue}}}
\renewcommand\subsection{%
  \@startsection{subsubsection}{2}{0pt}%
  {-\baselineskip}{.1\baselineskip}%
  {\normalfont\normalsize\bfseries\raggedright}}

\renewcommand\paragraph{%
  \@startsection{paragraph}{4}{0pt}%
  {3.25ex\@plus 1ex\@minus .2ex}{-1em}%
  {\normalfont\normalsize\bfseries}}
\makeatother

\newcommand{\wc}{{}\cdot{}}
\newcommand{\TM}[1]{\text{\textsc{#1}}}
\newcommand{\UTM}{\TM{UTM}}

\newcommand{\hrow}{h_{\text{row}}}
\newcommand{\hcol}{h_{\text{col}}}

\usepackage[textwidth=1in,textsize=scriptsize,colorinlistoftodos]{todonotes}

\usepackage{comment}
\includecomment{nowordcount}

\begin{document}

\begin{nowordcount}
\title{Undecidability of the Spectral Gap}

\author{Toby S.~Cubitt}
\affiliation{Department of Computer Science, University College London,
  Gower Street, London WC1E 6BT, United Kingdom}
\affiliation{DAMTP, University of Cambridge,
  Centre for Mathematical Sciences,
  Wilberforce Road, Cambridge CB3 0WA, United Kingdom}
\author{David Perez-Garcia}
\affiliation{Departamento de An\'alisis Matem\'atico and IMI,
  Facultad de CC Matem\'aticas,
  Universidad Complutense de Madrid,
  Plaza de Ciencias 3, 28040 Madrid, Spain}
\affiliation{ICMAT, C/Nicol\'as Cabrera, Campus de Cantoblanco,
  28049 Madrid, Spain}
\author{Michael M.~Wolf}
\affiliation{Department of Mathematics, Technische Universit\"at M\"unchen,
  85748 Garching, Germany}
\end{nowordcount}



\makeatletter
\let\frontmatter@abstractwidth\linewidth
\makeatother

\begin{nowordcount}
\begin{abstract}
  {\bf%
    The spectral gap---the energy difference between the ground state and first excited state---is central to quantum many-body physics. Many challenging open problems, such as the Haldane conjecture, existence of gapped topological spin liquid phases, or the Yang-Mills gap conjecture, concern spectral gaps. These and other problems are particular cases of the general spectral gap problem: given a quantum many-body Hamiltonian, is it gapped or gapless?
    Here we prove that this is an undecidable problem. We construct families of quantum spin systems on a 2D lattice with translationally-invariant, nearest-neighbour interactions for which the spectral gap problem is undecidable. This result extends to undecidability of other low energy properties, such as existence of algebraically decaying ground-state correlations.
    The proof combines Hamiltonian complexity techniques with aperiodic tilings, to construct a Hamiltonian whose ground state encodes the evolution of a quantum phase-estimation algorithm followed by a universal Turing Machine. The spectral gap depends on the outcome of the corresponding Halting Problem.
    Our result implies that there exists no algorithm to determine whether an arbitrary model is gapped or gapless. It also implies that there exist models for which the presence or absence of a spectral gap is independent of the axioms of mathematics.
  }
\end{abstract}
\end{nowordcount}

\maketitle

The spectral gap is one of the most important physical properties of a quantum many-body system, determining much of its low-energy physics. Gapped systems exhibit non-critical behaviour (e.g.\ massive excitations, short-range correlations). Whereas phase transitions occur when the spectral gap vanishes and the system exhibits critical behaviour (e.g.\ massless excitations, long-range correlations).
Many seminal results in condensed matter theory prove that specific systems are gapped or gapless. For example, Lieb, Schultz and Mattis' proof that the Heisenberg chain is gapless for half-integer spin\cite{LiebSchultzMattis} (later extended to higher dimensions by Hastings\cite{Hastings_LSM}), or Affleck et al.'s proof that the 1D AKLT model is gapped\cite{AKLT}. Similarly, many famous and long-standing open problems in theoretical physics concern the presence or absence of a spectral gap. A paradigmatic example is the antiferromagnetic Heisenberg model in 1D with integer spins. The ``Haldane conjecture'' that this model is gapped, first formulated in 1983\cite{Haldane}, has so far resisted rigorous proof despite strong supporting numerical evidence\cite{Haldane_numerics}. The same question in the case of 2D non-bipartite lattices such as the Kagome lattice was posed by Anderson in 1973\cite{Anderson73}. The latest numerical evidence\cite{White} strongly indicates that these systems may be topological spin liquids. This problem has attracted significant attention recently\cite{Balents} as materials such as herbertsmithite \cite{Han} have emerged whose interactions are well-approximated by the Heisenberg coupling. The presence of a spectral gap in these models remains one of the main unsolved questions concerning the long-sought topological spin liquid phase. In the related setting of quantum field theory, one of the most notorious open problems again concerns a spectral gap: the Yang-Mills mass gap problem\cite{Yang-Mills}. Proving a gap could provide a full explanation of the phenomenon of quark confinement. Whilst there is strong supporting evidence from lattice QCD numerics\cite{lattice_QCD}, the problem remains open.

All of these problems are specific instances of the general spectral gap problem: Given a quantum many-body Hamiltonian, is the system it describes gapped or gapless?
Our main result is to prove that the spectral gap problem is undecidable in general. This says more than merely showing that the problem is computationally or mathematically hard. Though one may be able to solve the spectral gap problem in specific cases, our result implies that it is in general logically impossible to say whether a system is gapped or gapless. This has two meanings, and we prove both:
\begin{enumerate}[label=\arabic*.,ref=\arabic*,leftmargin=*,labelsep=1em]
\item The spectral gap problem is \emph{algorithmically undecidable}: there cannot exist any algorithm which, given a description of the local interactions, determines whether the resulting model is gapped or gapless. This is the same sense in which the Halting Problem is undecidable\cite{Turing}.
\item The spectral gap problem is \emph{axiomatically independent}: given any consistent recursive axiomatisation of mathematics, there exist particular quantum many-body Hamiltonians for which the presence or absence of the spectral gap is not determined by the axioms of mathematics. This is the form of undecidability encountered in G\"odel's incompleteness theorem\cite{Godel}.
\end{enumerate}

\section*{Precise statement of results}
It is important to be precise in what we mean by the spectral gap problem. To this end, we must first specify the systems we are considering. Since we are proving undecidability, the simpler the system, the stronger the result. We restrict ourselves to nearest-neighbour, translationally-invariant spin lattice models on a 2D square lattice of size $L\times L$ (which we later take to $\infty$), with local Hilbert space dimension $d$. Any such Hamiltonian $H_L$ is completely specified by at most three finite-dimensional Hermitian matrices describing the local interactions of the system: two $d^2\times d^2$ matrices $\hrow$ and $\hcol$ specifying the interactions along the rows and columns, and a $d\times d$ matrix $h_1$ specifying any on-site interaction. All matrix elements will be algebraic numbers, and we normalise the interaction strength $\max\{\norm{\hrow},\norm{\hcol},\norm{h_1}\}=1$.

We must also be precise in what we mean by ``gapped'' and ``gapless'' (see \cref{fig:gapped-gapless}). Since quantum phase transitions occur in the thermodynamic limit of arbitrarily large system size, we are interested in the spectral gap $\Delta(H_L) = \lambda_1(H_L)-\lambda_0(H_L)$ as the system size $L\to\infty$ (where $\lambda_0,\lambda_1$ are the lowest and second-lowest eigenvalues). We take ``gapped'' to mean the system has a \emph{unique} ground state and a \emph{constant} lower bound on the spectral gap: $\Delta(H_L) \geq \gamma>0$ for all sufficiently large $L$. We take ``gapless'' to mean the system has \emph{continuous} spectrum above the ground state in the thermodynamic limit.

\begin{nowordcount}
\begin{figure}
  \includegraphics[scale=.47]{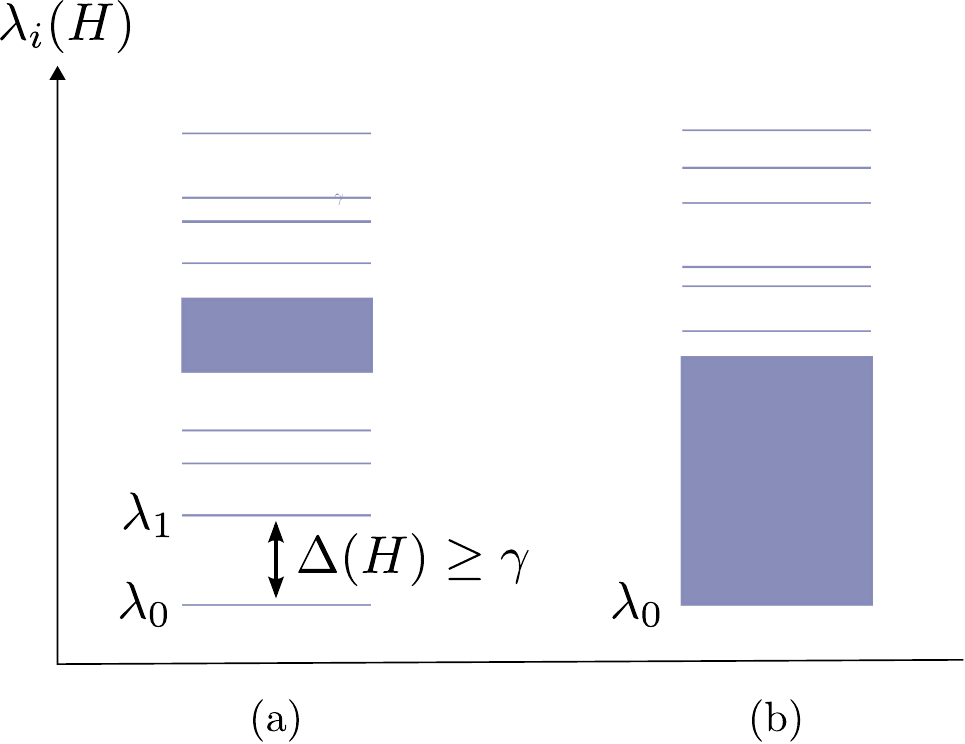}
  \caption[Gapped and gapless systems]{Gapped and gapless systems. (a)~A gapped system has a unique ground state $\lambda_0(H)$ and a constant lower-bound $\gamma$ on the spectral gap $\Delta(H)$ in the thermodynamic limit. (b)~A gapless system has continuous spectrum above the ground state in the thermodynamic limit.}
  \label{fig:gapped-gapless}
\end{figure}
\end{nowordcount}

Note that gapped is not the negation of gapless here; there are systems that fall into neither category. We adopt such strong definitions to deliberately exclude ambiguous cases, such as systems with degenerate ground states. A Hamiltonian that is gapped or gapless according to the above definitions is recognised as such throughout the literature. We show that the spectral gap problem is undecidable even with the promise that the Hamiltonian \emph{either} has a unique ground state and a spectral gap of magnitude~1, \emph{or} has continuous spectrum above the ground state.

We prove this by showing that the Halting Problem for Turing Machines can be encoded in the spectral gap problem, implying that the latter is at least as hard as the former. Recall that a Turing Machine is a simple, abstract model of computation in which a head reads and writes symbols from some finite alphabet on an infinite tape and moves left or right, following a finite set of rules. The Halting Problem asks: given an initial input written on the tape, does the Turing Machine halt on that input? Turing famously proved that this problem is undecidable\cite{Turing}, and we relate it to the spectral gap problem in the following way:

\begin{theorem}\label{thm:promise}
  We can construct explicitly a dimension $d$, $d^2\times d^2$ matrices $A,B,C,D$ and a rational number $\beta>0$, which can be chosen as small as desired, so that:
  \begin{enumerate}
  \item $A$ is Hermitian, with matrix elements in $\Z+\beta \Z+ \frac{\beta}{\sqrt{2}}\Z$,
  \item $B,C$ have integer matrix elements,
  \item $D$ is Hermitian, with matrix elements in $\{0,1,\beta\}$.
  \end{enumerate}
  For each positive integer $n$, define the local interactions of a translationally-invariant, nearest-neighbour Hamiltonian $H(n)$ on a 2D square lattice as:
  \begin{gather*}
    h_1(n)=\alpha(n)\Pi\\
    h_{\text{row}}=D\\
    h_{\text{col}}(n)=A + \beta\left(e^{i\pi\varphi(n)} B + e^{-i\pi\varphi(n)} B^\dg + e^{i\pi2^{-\abs{\varphi(n)}}} C + e^{-i\pi2^{-\abs{\varphi(n)}}} C^\dg\right),
  \end{gather*}
  ($\varphi(n) = n/2^{\abs{n}-1}$ is the rational number whose binary fraction expansion contains the binary digits of $n$ after the decimal point, $\abs{\varphi}$ denotes the number of digits in this expansion, $\alpha(n)$ is an algebraic number $\le \beta$ computable from $n$, and $\Pi$ is a projector.) Then:
  \begin{enumerate}
  \item The local interaction strength is $\le 1$ (i.e.\ $\norm{h_1(n)}$, $\norm{h_{\text{row}}}$, $\norm{h_{\text{col}}(n)} \leq 1$).
  \item If the universal Turing Machine halts on input $n$, then the Hamiltonian $H(n)$ is gapped with $\gamma\ge 1$.
  \item If the universal Turing Machine does not halt on input $n$, then the Hamiltonian $H(n)$ is gapless (i.e.\ has continuous spectrum).
  \end{enumerate}
\end{theorem}
This implies that the spectral gap problem is algorithmically undecidable since the Halting Problem is. By a standard argument \cite{Poonen} this also implies axiomatic independence. Both forms of undecidability extend to other low-temperature properties, such as critical correlations in the ground state. In fact, our method allows us to prove undecidability of \emph{any} physical property that distinguishes a Hamiltonian from a gapped system with unique, product ground state.


\section*{Hamiltonian construction}
We will first relate undecidability of the spectral gap to undecidability of another important physical quantity, the ground state energy density, which for a 2D lattice is given by $E_\rho = \lim_{L\to\infty} \lambda_0(H_L)/L^2$. We then show how to transform the Halting Problem into a question about ground state energy densities.

Reducing the ground state energy density problem to the spectral gap problem requires two ingredients:
\begin{enumerate}
\item \label{h_u}
  A translationally-invariant Hamiltonian $H_u(\varphi)$ on a 2D square lattice with local interactions $h_u(\varphi)$, whose ground state energy density is either strictly positive or tends to $0$ from below in the thermodynamic limit, depending on the value of an external parameter $\varphi$. However, determining which case holds should be undecidable. Constructing such a Hamiltonian constitutes the main technical work of our result. (Note that these properties are unaffected if we multiply $h_u(\varphi)$ by an arbitrary fixed rational number $\beta$, no matter how small.)
\item \label{h_d}
  A gapless Hamiltonian $H_d$ with translationally-invariant local interactions $h_d$ and ground state energy~0.\footnote{Recall from above that by ``gapless'' we mean continuous spectrum above the ground state, not merely a vanishing spectral gap.} There are many well-known examples of such Hamiltonians, e.g.\ the critical XY-model\cite{LiebSchultzMattis}.
\end{enumerate}
Given Hamiltonians with these properties, we construct a new translationally-invariant Hamiltonian with local interactions $h(\varphi)$ which is gapped or gapless, depending on the value of $\varphi$, in the following way. The local Hilbert space of $h(\varphi)$ is the tensor product of those of $h_u$ and $h_d$ together with one additional energy level: $\HS = \ket{0}\oplus\HS_u\ox\HS_d$. We then take the interaction $h^{(i,j)}$ between nearest-neighbour sites $i$ and $j$ to be:
\begin{equation}\label{eq:promise_H}
  h(\varphi)^{(i,j)}
    = \proj{0}^{(i)} \ox (\1-\proj{0})^{(j)}
      + h_{\vphantom{d}u}^{(i,j)}(\varphi)\ox \1_d^{(i,j)}
      + \1_{\vphantom{d}u}^{(i,j)}\ox h_d^{(i,j)}.
\end{equation}
Its spectrum can be seen to be (see supplementary information for details):
\begin{equation}\label{spectrum}
  \spec H(\varphi) = \{0\}\cup\Big\{\spec H_u(\varphi) + \spec H_d\Big\}\cup S,
\end{equation}
with $S\ge 1$. Recalling that we chose $H_d$ to be gapless, we see immediately from \cref{spectrum} that if the ground state energy density of $H_u$ tends to zero from below (so that $\lambda_0(H_u) < 0$) then $H(\varphi)$ is gapless, but if $H_u$ has strictly positive ground state energy density (so that $\lambda_0(H_u)$ diverges to $+\infty$) then it has spectral gap $\geq 1$, as required (see \cref{fig:GSE-to-gap}).

%
Observe that this construction is rather general: by choosing different $h_d$ we obtain undecidability of \emph{any} physical property that distinguishes a Hamiltonian from a gapped system with unique product ground state.

\begin{nowordcount}
\begin{figure}
  \includegraphics[scale=.47]{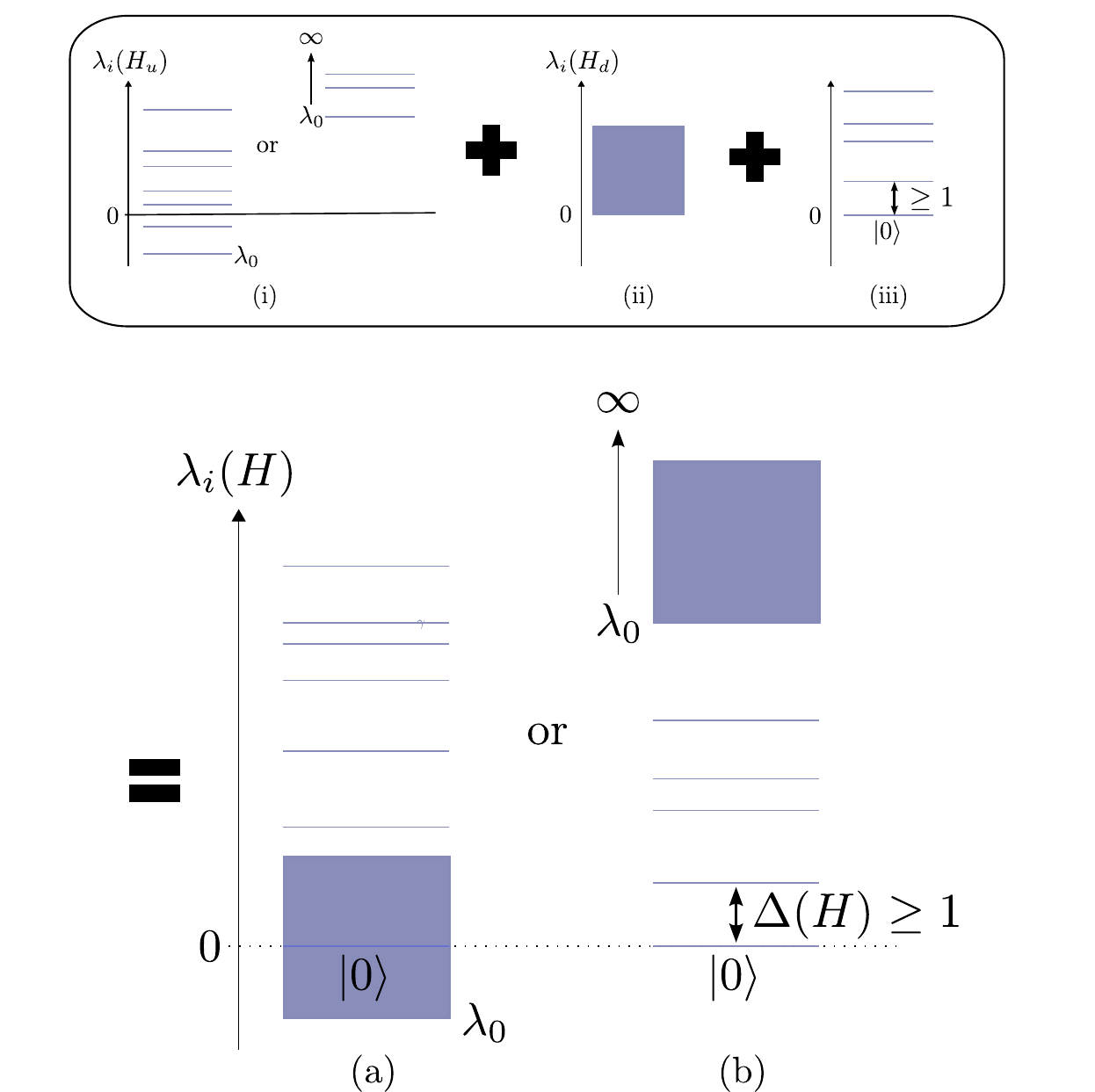}
  \caption[Ground state energy density to spectral gap.]{Ground state energy density to spectral gap. To relate ground state energy density and spectral gap, we need: (i)~a Hamiltonian $H_u(\varphi)$ whose ground state energy density is either strictly positive or tends to $0$ from below in the thermodynamic limit, but determining which is undecidable, and (ii)~a gapless Hamiltonian $H_d$ with ground state energy~0. We combine $H_u(\varphi)$ and $H_d$ to form a new local interaction, $h(\varphi)$, in such a way that $h(\varphi)$ has (iii)~an additional non-degenerate 0-energy eigenstate, and the continuous spectrum of $H_d$ is shifted immediately above the ground state energy of $H_u$.
  %
  (a)~If the ground state energy density of $H_u(\varphi)$ tends to~0 from below, then its ground state energy in the thermodynamic limit must be $\leq 0$, and $h(\varphi)$ is gapless.
  (b)~Whereas if the ground state energy density of $H_u(\varphi)$ is strictly positive, its ground state energy in the thermodynamic limit must diverge to $+\infty$, and $h(\varphi)$ is gapped.}
  \label{fig:GSE-to-gap}
\end{figure}
\end{nowordcount}

\subsection*{Encoding computation in ground states}
To construct the Hamiltonian $H_u(\varphi)$, we encode the Halting Problem into the local interactions $h_u(\varphi)$ of the Hamiltonian. The latter concerns the dynamics of a classical system---a Turing Machine. How can this be related to the ground state energy density---a static property of a quantum system? The idea, which dates back to Feynman\cite{Feynman}, is to construct a Hamiltonian whose ground state encodes the entire history of the computation in superposition. I.e.\ if the state of the computation at time $t$ is represented by the state vector $\ket{\psi_t}$, the ground state is the so-called \emph{computational history state} $\tfrac{1}{\sqrt{T}}\sum_{t=0}^{T-1}\ket{t}\ket{\psi_t}$. In the following, we will sometimes refer to the Turing Machines encoded in the Hamiltonian ``running'' on some input. This should be understood to mean that the evolution produced by running the Turing Machine on that input appears in the ground state as the corresponding computational history state.

If there are no other constraints, writing down such a Hamiltonian is straightforward. However, constructing such a Hamiltonian out of the local interactions of a many-body system is more involved. Feynman's idea was substantially developed by Kitaev\cite{Kitaev_book}, and after a long sequence of results\cite{KKR,Oliveira-Terhal,AGIK} culminated in Gottesman and Irani's construction for 1D spin chains with translationally-invariant, nearest-neighbour interactions\cite{Gottesman-Irani}.

For any Quantum Turing Machine\cite{Bernstein-Vazirani} (QTM), Gottesman and Irani construct an interaction $h$ between neighbouring particles, such that the ground state of the 1D translationally-invariant Hamiltonian $H_{\text{GI}} = \sum_{i=1}^N h_{i,i+1}$ is of the form $\tfrac{1}{\sqrt{T}}\sum_{t=0}^{T-1}\ket{t}\ket{\psi_t}$, where the ``clock'' part of the computational history state $\ket{t} \simeq \ket{1}^{\ox t}\ket{0}^{\ox N-t}$ counts time in unary, and $\ket{\psi_t}$ represents the state of the QTM after $t$ time-steps. Moreover, the ground state energy can be taken equal to $0$.

The translationally-invariant Hamiltonians we are considering are completely specified by the finite number of matrix elements in the local interactions $\hrow$, $\hcol$ and $h_1$. To encode the Halting Problem in the Hamiltonian, we need a way of encoding any of the countably infinite possible inputs to the universal Turing Machine into these parameters. For these, we make use of quantum phase estimation\cite{Nielsen+Chuang}.

\subsection*{Quantum phase estimation}
Given a unitary $U$, the quantum phase estimation algorithm estimates an eigenvalue $e^{2\pi i\varphi}$ of $U$ to a given number of bits of precision (which must be chosen in advance). It is well-known\cite{Nielsen+Chuang} that if the number of bits of precision in the quantum phase estimation algorithm is greater than or equal to the number of digits in the binary fraction expansion of $\varphi$, then the quantum phase estimation algorithm, rather than estimating the phase approximately, will output all the digits of $\varphi$ (written as a binary fraction) exactly.

We use this property to construct a family of Quantum Turing Machines $P_n$, indexed by $n\in\N$, with the following properties:
\begin{inparaenum}
\item The number of internal states and tape symbols of $P_n$ are independent of $n$.
\item Given a number $N = 2^x-1 \geq n$ as input (written in binary), $P_n$ writes the binary expansion of $n$ on its tape and then halts deterministically. (The reason for having input $N$ of this form will become clear later).
\end{inparaenum}
To construct $P_n$, we construct a QTM which uses the input $N$ to determine how many digits of precision to use, then runs the quantum phase estimation algorithm on the single-qubit gate $U = \left(\begin{smallmatrix} 1 & 0\\ 0 & e^{2\pi i\varphi} \end{smallmatrix}\right)$. The phase $\varphi$ in $U$ is determined by the transition rules of the QTM\cite{Bernstein-Vazirani}. Choosing $\varphi$ to be the rational number whose binary fraction expansion contains the digits of $n$ (expressed in binary) achieves the desired behaviour for $P_n$. By ``dovetailing'' $P_n$ with a universal Turing Machine $\UTM$ (i.e.\ run $P_n$ first, then run $\UTM$), the UTM runs on the input specified by $\varphi$.

The quantum computation carried out by $P_n$ followed by $\UTM$ can be encoded in the Hamiltonian using the history state construction described above. The phase $e^{2\pi i\varphi}$ being estimated then becomes one of the matrix elements of the Hamiltonian. The same happens with the $e^{i\pi2^{-\abs{\varphi}}}$ appearing in the inverse Quantum Fourier Transform---the key ingredient of the quantum phase estimation algorithm.

Finally, we must ensure that the $\ket{\psi_0}$ component of the history state is correctly initialised to input of the form $N=2^x-1$ (written in binary) required by $P_n$. But $N=2^x-1$ in binary is simply a string of $N$ `1's, and it is easy to ensure that $\ket{\psi_0}$ is the state $\ket{1}^{\ox N}$ using translationally-invariant local interactions.



If we add an on-site interaction $h_1=\proj{\top}$ to the history state Hamiltonian constructed above, giving an additional energy to the state $\ket{\top}$ representing the halting state, then its ground state will pick up additional energy if and only if the UTM halts. However, the ground state energy still converges to $0$ with system size in both cases (see supplementary information). The energy \emph{density} therefore tends to zero in the thermodynamic limit, whether or not the UTM halts.

To remedy this, and amplify the difference between the halting and non-halting cases, we use the second spatial dimension and exploit \emph{Wang tilings}.

\subsection*{Quasi-periodic tilings}
A Wang tile\cite{Wang} is a square with markings along each edge. A tiling is then an arrangement of such tiles covering the whole plane, so that the markings on adjacent edges match.
A tiling can easily be encoded in a ground state of a \emph{classical} Hamiltonian on a 2D square lattice: by representing tile types by an orthogonal basis $\{\ket{T}\}$ for the local Hilbert space $\HS_c$, and choosing local interaction terms $\proj{T_i}\ox\proj{T_j}$ to give an energy penalty to all adjacent non-matching pairs of tiles $T_i,T_j$, a tiling of the plane is equivalent to a ground state with energy zero.


We prove and exploit very particular properties of an aperiodic tiling due to Robinson\cite{Robinson}, and combine this with the history state Hamiltonian. Although the pattern of tiles in the Robinson tiling extends infinitely in all directions, it never repeats. More precisely, it contains periodically repeating sub-patterns forming squares with sizes given by every power of~4 (see \cref{fig:Robinson_tiling+QTM}). This will allow us to encode in the ground state many copies of the UTM running on the same input $\varphi$,
with tapes of \emph{all} possible finite lengths and for every possible finite run-time (see \cref{fig:Robinson_tiling+QTM}).

This is achieved by sandwiching the 1D quantum history-state Hamiltonian $h_q$ ``on top of'' the Robinson tiling Hamiltonian $h_c$ to form two ``layers'', so that the local Hilbert space at each site is $\HS = \HS_c\ox(\HS_e\oplus\HS_q)$ (where $\HS_e = \ket{0}$ is an additional energy level). One can then construct a Hamiltonian (see supplementary information) whose ground state is of the form $\ket{T}_c\otimes \ket{\psi}_{eq}$, where $\ket{T}_c$ is a product state representing a classical configuration of the tiling layer and $\ket{\psi}_q$ contains -- in a tensor product structure -- computational history states along one edge (called a ``segment'') of all of the squares appearing in the configuration given by $T$. These computational history states are essentially the only constituents of $\ket{\psi}_q$ which contribute to the energy. The Hamiltonian also has an on-site interaction $h_1=\proj{\top}$ that gives an additional energy to the state $\ket{\top}$ representing the halting state of the Turing Machine. Hence the ground state will pick up additional energy from all encoded Turing Machines that halt. This energy still decreases with the effective size of the system which, however, is now the size of the corresponding segment in the Robinson tiling (see \cref{fig:Robinson_tiling+QTM}), \emph{not} the overall system size.

\begin{nowordcount}
\begin{figure}
  \includegraphics[scale=.64]{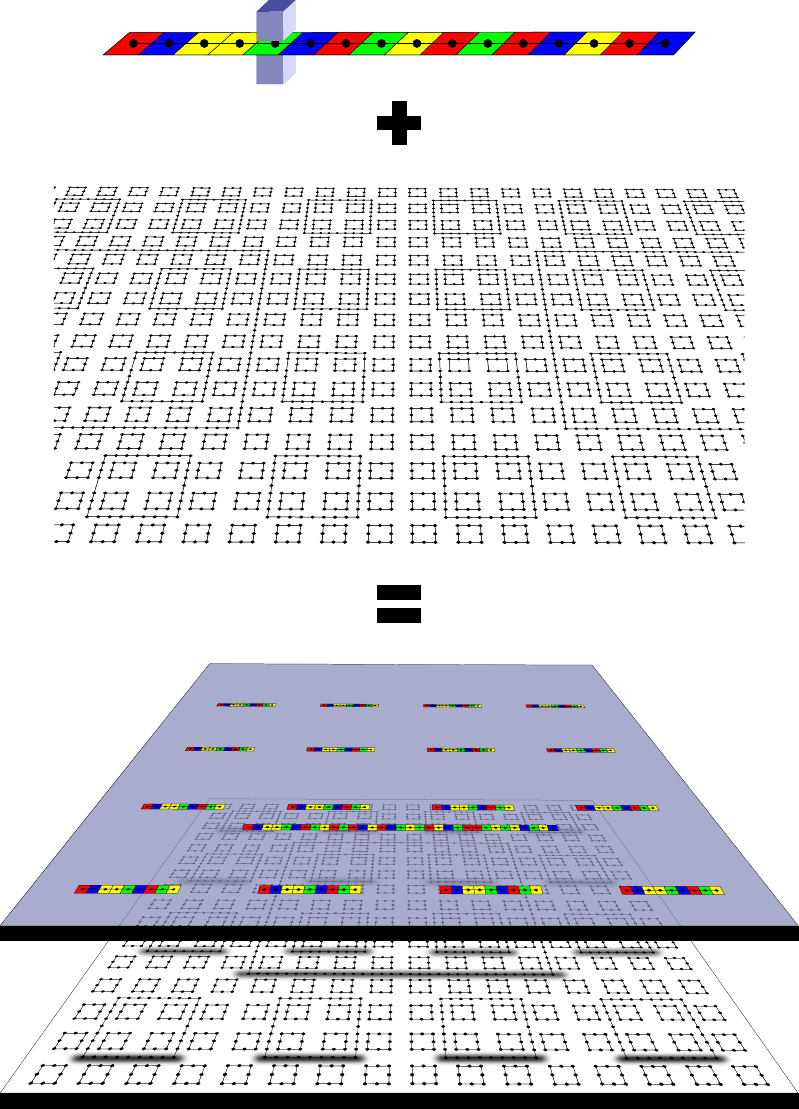}
  \caption[Complete Hamiltonian construction]{Complete Hamiltonian construction. The Robinson tiles enforce a recursive pattern of interlocking squares, of sizes given by all integer powers of~4 (middle figure). As with any Wang tiling, we can readily represent this as a classical Hamiltonian whose ground state has the same quasi-periodic structure. Since the set of tiles is fixed, the local dimension of this Hamiltonian is constant. By adding a ``quantum layer'' on top of this tiling Hamiltonian and choosing a suitable translationally-invariant coupling between the layers, we can effectively place copies of QTM encoded in a 1D history state Hamiltonian (top figure) along one edge of all of the squares. The ground state of this Hamiltonian consists of the Robinson tiling configuration in the tiling layer, with computational history states in the quantum layer along one edge of all of each square in the tiling (bottom figure). Each of these encodes the evolution of the \emph{same} quantum phase estimation algorithm and universal Turing Machine. The effective tape length available each QTM is determined by the size of the square it ``runs'' on.}
  \label{fig:Robinson_tiling+QTM}
\end{figure}
\end{nowordcount}



What can we say about the ground state energy? If the UTM does not halt on input $n$, then $\ket{T}_c$ is a valid tiling and for all segments which are larger than $\abs{n}$ the ground state energy contribution is $0$. The contribution for each segment smaller than $\abs{n}$ is some algebraic computable number. If $\alpha(n)$ is the constant that adds up the contributions of all segments smaller then $\abs{n}$, the addition of the energy shift $h_1 = -\alpha(n)\id$ to the Hamiltonian makes the ground state energy density negative (and tending to~0 from below) in the non-halting case (see supplementary information).

In the halting case, one of two things may happen. If $\ket{T}_c$ is a valid tiling, the number of squares large enough for the encoded Turing Machine to halt grows quadratically with system size, and each of them contributes a small but non-zero energy. Since such a state also picks up the energy contribution from segments of size smaller than $\abs{n}$, even after adding $h_1 = -\alpha(n)\id$ the energy diverges with lattice size. Hence the ground state energy density is strictly positive in the halting case, as desired.

Alternatively, one could try to reduce the energy by introducing defects in the tiling, that effectively ``break'' some of the Turing Machines so that they do not halt. However, we prove that the Robinson tiling is robust against such defects: a tile mismatch can only affect the pattern of squares in a finite region around the defect,
and each defect contributes $O(1)$ energy. We can choose the parameters (see supplementary information) to guarantee that introducing defects is energetically unfavourable. This completes the argument for our main result. Additional technical details can be found in the supplementary information (online).

\section*{Discussion}
Let us take a step back and discuss both the implications and the limitations of this result. The result clearly concerns mathematical models of quantum many-body systems, but also the behaviour of and methods for treating the thermodynamic limit. Moreover, the result can also be seen as an indication of new physical phenomena.

An immediate consequence of our result is that there cannot exist an algorithm or a computable criterion that solves the spectral gap problem in general. Whilst algorithmic undecidability always concerns infinite families of systems, the axiomatic interpretation of the result also allows us to talk about individual systems: there are particular Hamiltonians within these families for which one can neither prove nor disprove the presence of a gap, or of any other undecidable property. Unfortunately, our methods cannot pinpoint these cases, let alone prove that one of the big open problems mentioned in the introduction is of this kind.

A further consequence concerns the behaviour of the thermodynamic limit. In practice, we usually probe the idealised infinite thermodynamic limit by studying how the system behaves as we take finite systems of increasing size. One often assumes that the systems, though finite, are so large that we already see the asymptotic behaviour. In numerical simulations of condensed matter systems, one typically simulates finite systems of increasing size and extrapolates the asymptotic behaviour from finite size scaling\cite{finite_size_scaling}. Similarly, lattice QCD calculations simulate finite lattice spacings, and extrapolate the results to the continuum\cite{lattice_QCD}. Renormalisation group techniques accomplish something similar mathematically\cite{Cardy_book}. However, the undecidable quantum many-body models constructed in this work exhibit behaviour that defeats such approaches. As the system size increases, the Hamiltonian will initially look like a gapless system, with the low-energy spectrum appearing to converge to a continuum. But at some threshold lattice size, a spectral gap of magnitude~1 will suddenly appear (or, vice versa, a gap will suddenly close\cite{long-version}). Not only can the lattice size at which the system switches from gapless to gapped be arbitrarily large, the threshold at which this transition occurs is uncomputable. The analogous implication also holds for all other undecidable low-temperature properties. Thus any method of extrapolating the asymptotic behaviour from finite system sizes must fail in general.

This leads us directly to new physical phenomena. First, it hints at a new type of ``phase transition'', not driven by temperature or extrinsic local parameters, but by the size of the system. Some of the models constructed in the proof exhibit a drastic and abrupt change of properties when their size is increased beyond a certain scale. The scale at which this happens can be very large and is, in fact, not generally computable from the local description of the system. 
Second, our results show that certain quantum many-body models exhibit a radical form of instability. An arbitrarily small change in the parameters can make the system cross an arbitrary number of gapped/gapless transitions. In a sense, this phenomenon is the source of the undecidability in our models.

We finish this discussion with a closer look at some of the limitations of the result. First, our results concern 2D (or higher-dimensional) systems. Although the majority of our construction is already 1D, we do not currently know whether the entire result holds in 1D as well. Second, although a theoretical model of a quantum many-body system is always an idealisation of the real physics, the models we construct in the proof are highly artificial. Whether the results can be extended to more natural models is open. A related point is that we prove our results for Hamiltonians with a very particular form. We do not know how stable the results are to small deviations from this. This is a general issue with most many-body models; stability in this sense is not understood even for much simpler models such as the Ising model. Recent stability proofs only apply to certain types of frustration-free Hamiltonian\cite{BravyiHastingsMichalakis,Michalakis}. Indeed, our results restrict the extent to which such stability results can be generalised. Similarly, we do not know whether the results hold for systems with small local Hilbert space dimension. Whilst the dimension $d$ in \cref{thm:promise} is fixed and finite, providing an estimate for it would be cumbersome and certainly involve large exponentials. However, the steps in the proof described above are not tailored to minimising this dimension. Whether there is a non-trivial bound on the local Hilbert space dimension below which the spectral gap problem becomes decidable is an intriguing open question.

\begin{nowordcount}



\vspace{.5em}

\noindent \textbf{Supplementary Information} is linked to the online version of the paper at \url{www.nature.com/nature}.

\vspace{.5em}

\noindent \textbf{Acknowledgements}
TSC thanks IBM T.\ J.\ Watson Laboratory for their hospitality, and Charlie Bennett in particular for discussions about this work.

\noindent TSC, DPG and MMW thank the Isaac Newton Institute for Mathematical Sciences, Cambridge for their hospitality during the programme ``Mathematical Challenges in Quantum Information'', where part of this work was carried out.
TSC is supported by the Royal Society.
DPG acknowledges support from MINECO (grant MTM2011-26912 and PRI-PIMCHI-2011-1071), Comunidad de Madrid (grant QUITEMAD+-CM, ref. S2013/ICE-2801), and from the European Research Council (ERC) under the European Union's Horizon 2020 research and innovation programme (grant agreement No 648913).

\noindent This work was made possible through the support of grant \#48322 from the John Templeton Foundation. The opinions expressed in this publication are those of the authors and do not necessarily reflect the views of the John Templeton Foundation.

\vspace{.5em}

\noindent \textbf{Author Contributions} All authors contributed extensively to the paper.

\vspace{.5em}

\noindent \textbf{Author Information} Reprints and permissions information is available at \url{www.nature.com/reprints}. There are no competing financial interests. Correspondence and requests for materials should be addressed to TSC \texttt{t.cubitt@ucl.ac.uk}.
\end{nowordcount}

\end{document}